\documentclass{iau}
\usepackage{graphicx}
\usepackage{natbib,aas_macros}

\title[Low frequency observations of relics and halos] 
{Low frequency observations of radio\\ relics and halos
}

\author[Ruta Kale]   
{Ruta Kale$^1$}
\affiliation{$^1$National Centre for Radio Astrophysics, \\ 
Tata Institute of Fundamental Research, \\ 
S. P. Pune University Campus,\\
Ganeshkhind, Pune 411007, India \\ email: {\tt ruta@ncra.tifr.res.in 
}}

\pubyear{2018}
\volume{342}  
\pagerange{119--126}
\jname{Perseus in Sicily: from black hole to cluster outskirts}
\editors{Keiichi Asada, Elisabete de Gouveia dal Pino, et al}
\begin{document}

\maketitle

\begin{abstract}
Diffuse radio emission from galaxy clusters in the form of radio halos and relics are 
tracers of the shocks and turbulence in the intra-cluster medium. The imprints of the physical processes that govern their origin and evolution can be found in their radio morphologies and spectra. The role of mildly relativistic population of electrons may be crucial for the acceleration mechanisms to work efficiently. Low frequency observations with telescopes that allow imaging of extended sources over a broad range of low frequencies ($< 2$ GHz) offer the best tools to study these sources. I will review the Giant Metrewave Radio Telescope (GMRT) observations in the past few years that have led to: i) statistical studies of large samples of galaxy clusters, ii) opening of the discovery space in low mass clusters and iii) tracing the spectra of seed relativistic electrons using the Upgraded GMRT.
\keywords{acceleration of particles, radiative mechanisms: non-thermal, shock waves, turbulence, galaxies:clusters:general}
\end{abstract}

\firstsection 
\section{Introduction}

\begin{figure}[t]
\begin{center}
\includegraphics[trim= 0.0cm 0.0cm 0.0cm 0.0cm,clip,height=2.15in]{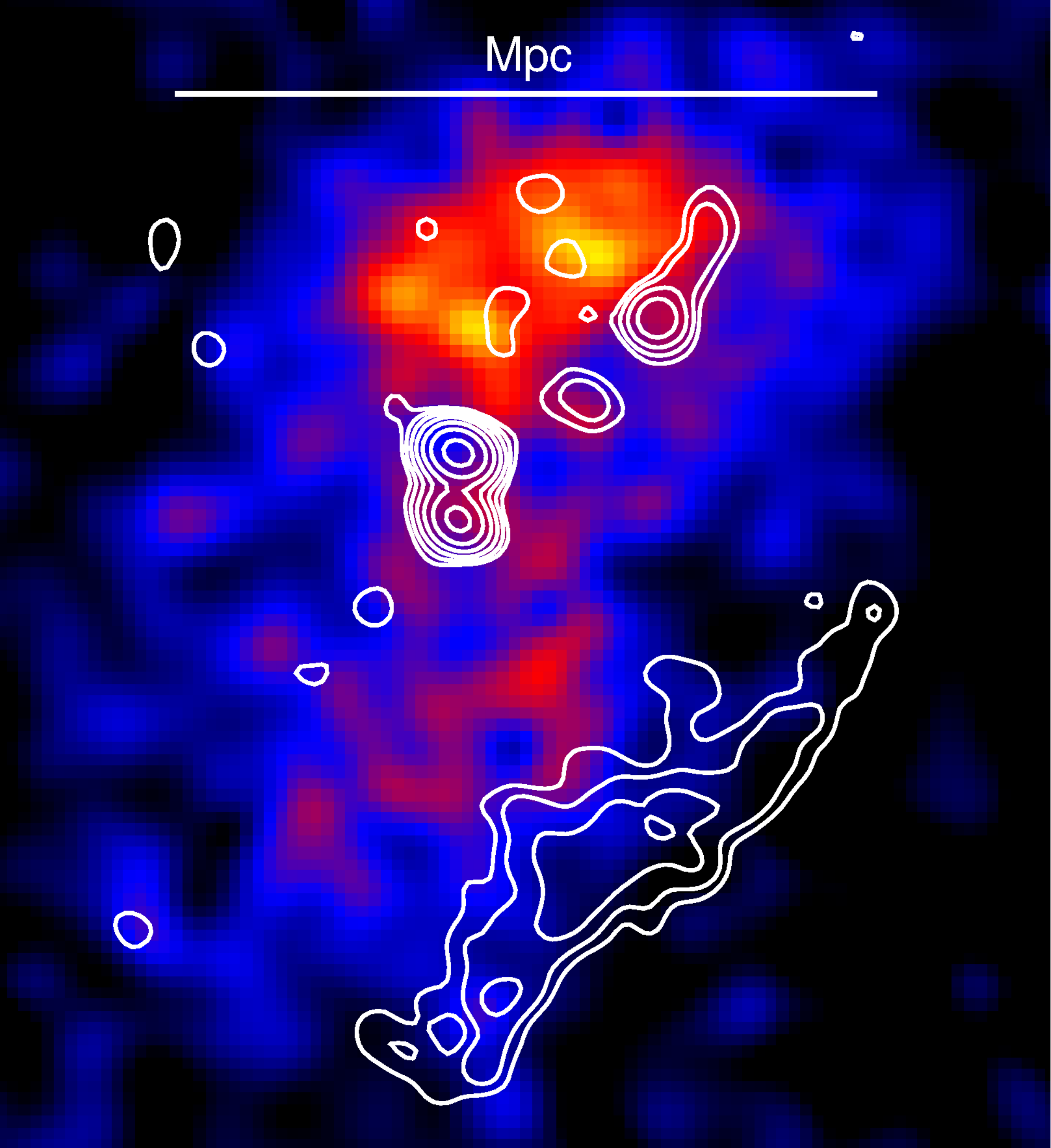}
\includegraphics[trim= 0.0cm 0.0cm 0.0cm 0.0cm,clip,height=2.15in]{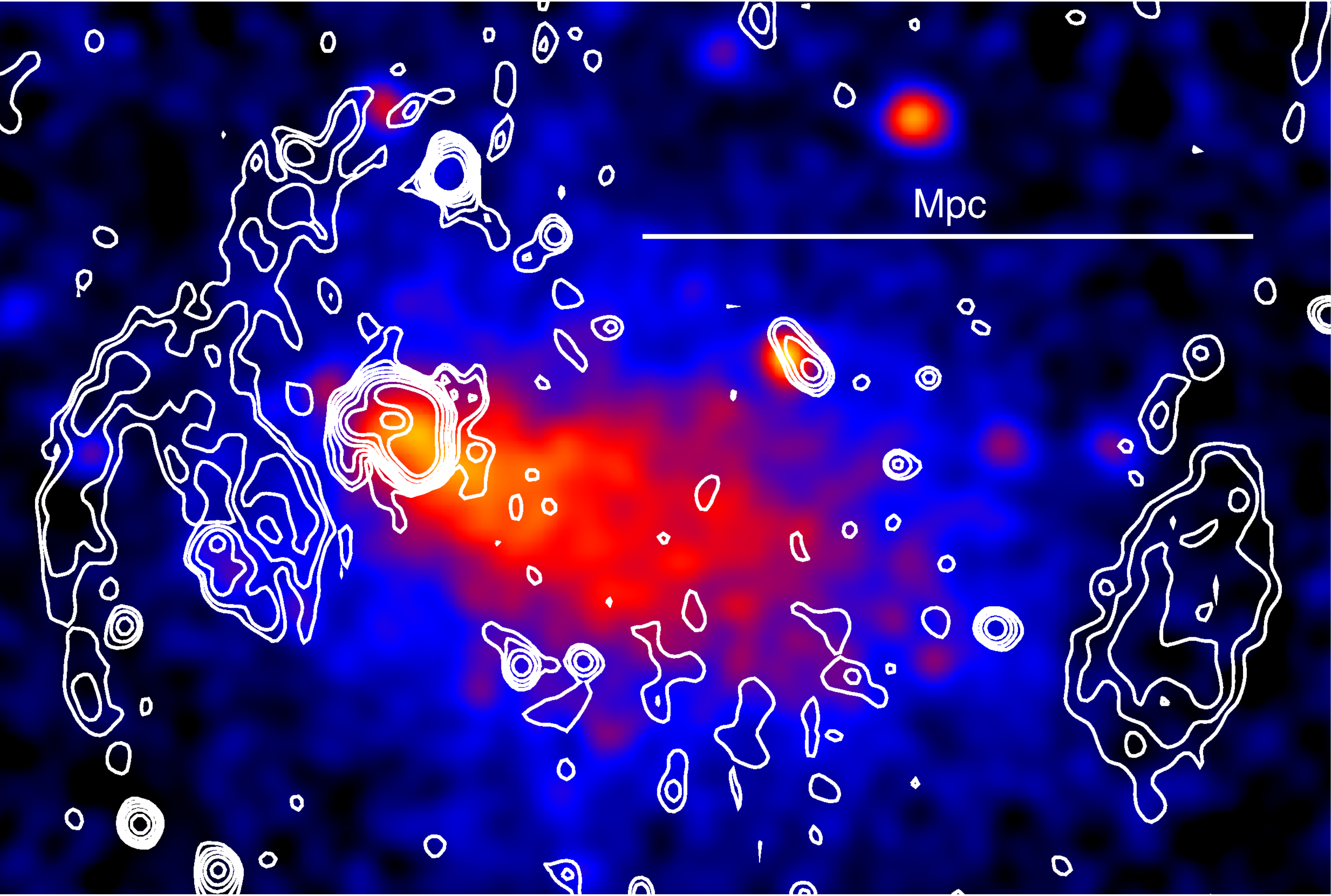} \\
\includegraphics[trim= 0.0cm 0.0cm 0.0cm 0.0cm,clip,height=2.0in]{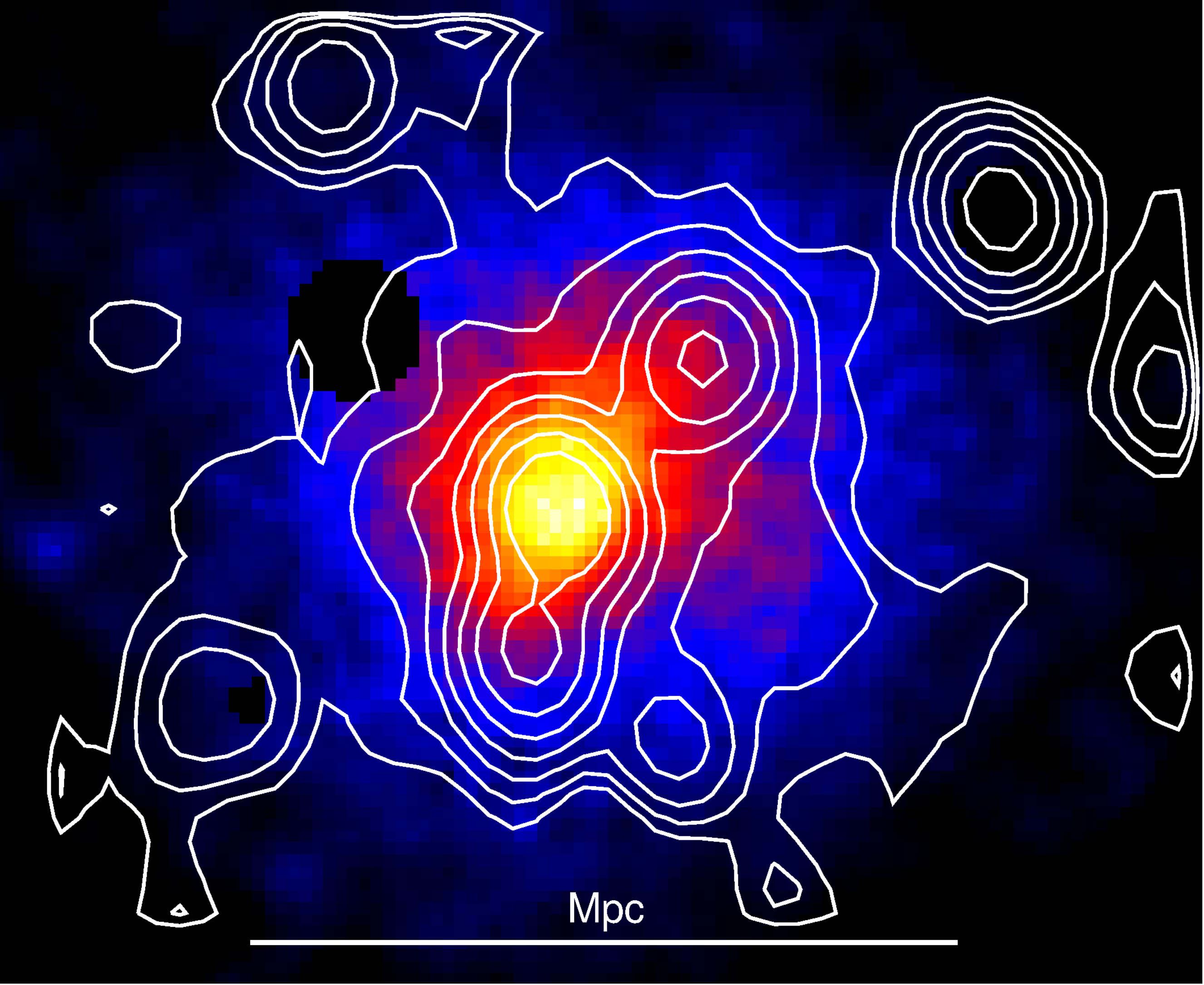} 
 \includegraphics[trim= 0.0cm 0.0cm 0.0cm 0.0cm,clip,height=2.0in]{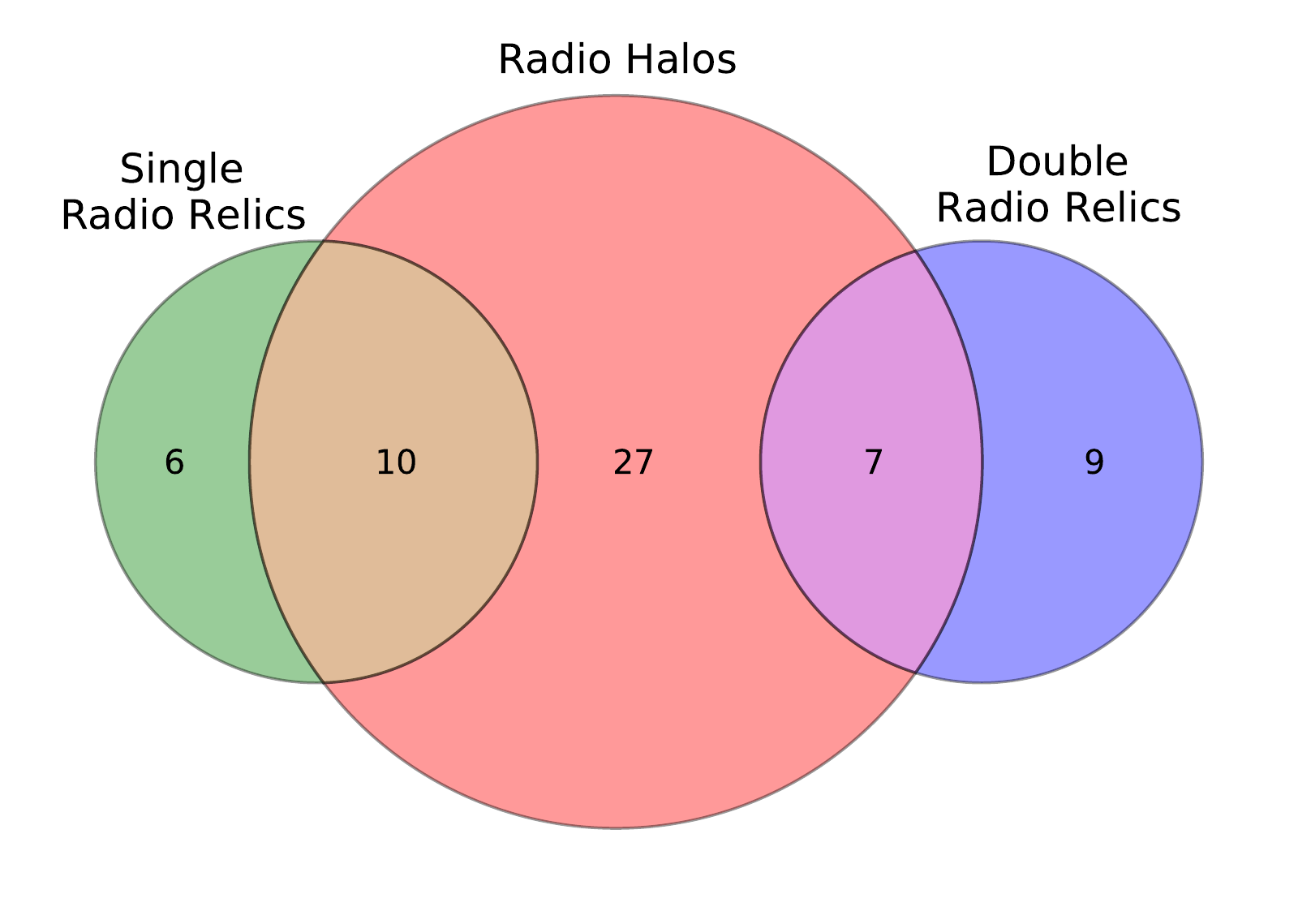} 
 \caption{The single radio relic in the cluster PLCK G200.9-28.2 (top left), the double radio relic in Abell 3376 (top right) and the radio halo in PLCKG171.9-40.7 (bottom left) are shown in white radio contours overlaid on the respective X-ray images of the host clusters in colour. The relics are elongated arc-like sources at the edges of the clusters and a radio halo is a centrally located Mpc-scale extended radio source. Bottom right:- The Venn diagram represents the currently known clusters that are host to one or more kinds of these emission. There are significant number of clusters that are host to one or more kinds of such radio emission. The sample is taken from  \citet{2015ApJ...813...77Y}.}
   \label{eg}
\end{center}
\end{figure}

Clusters of galaxies are gravitationally bound systems of masses $\sim 10^{14}-10^{15}$ M$_{\odot}$ that are composed of dark matter, galaxies and the intra-cluster medium (ICM). The ICM is the most massive baryonic component forming $10-15\%$ of the total mass and mainly contains thermal plasma that emits in X-rays via thermal Bremsstrahlung mechanism. It also contains non-thermal components such as the magnetic fields and relativistic particles but these elude detection in most observing bands.
The detection of diffuse radio emission of synchrotron origin from the ICM provides direct evidence for the presence of relativistic electrons ($\sim $GeV) and magnetic fields ($\sim 0.1 - $ a few $\mu$G) in galaxy clusters. These sources are typically classified into radio halos and radio relics based on their morphology and location relative to the X-ray emitting thermal ICM \citep[see][for reviews]{fer12, bru14}. Due to their steep spectra ($\alpha > 1.0$, $\rm S_{\nu} \propto \nu^{-\alpha}$) these sources are typically studied at low radio frequencies ($\leq 2 $GHz). The short radiative lifetime ($\sim 0.1$ Gyr) and long diffusion times ($\geq $Gyr to reach Mpc distance) of relativistic electrons in the ICM requires that radio halos and relics have mechanisms of in-situ re-acceleration associated with their origin \citep[e. g.][]{jaf77}. 

Radio relics are elongated or arc-like, polarized radio sources that are found at the peripheries of clusters. These occur as single or sometimes in pairs around galaxy clusters (Fig.~\ref{eg}, top). Radio relics are proposed to trace shocks at the cluster outskirts where particles are accelerated \citep[e.g.][]{ens98}. The polarization indicating aligned magnetic fields, spectral indices showing steepening from outer to the inner edges of the relics \citep[e. g.][]{gia08,bon09,wee10,kal12} and the co-spatiality with X-ray detected shocks \citep[e. g.][]{aka12,ogr13} provide support for relics as tracers of shocks.

Radio halos are centrally located in the cluster, Mpc sized and unpolarized (Fig.\ref{eg}, bottom left). The origin of radio halos had been proposed to be in the secondary electrons generated by the hadronic collisions in the ICM \citep[e. g.][]{den80,bla99}. The stringent upper limits on the gamma rays associated with this process \citep[e. g.][]{ack10,arlen12} have led to a scenario where a primary mechanism such as turbulent reacceleration \citep[e. g.][]{pet01,bru01} may be playing a crucial role \citep[e. g.][]{bru17}. An empirical scaling relation between the cluster mass (or X-ray luminosity) and the radio power of radio halos is known \citep[e. g.][]{cas13}. 

The connection between cluster mergers and occurrence of radio halos and relics has been found observationally \citep{buo01,cas10,kal15}. Cluster mergers are a natural origin for the shocks and turbulence that are proposed to play a role in the generation of such sources. Indeed a significant fraction of clusters that host radio halos and relics show presence of both types of sources (Fig.~\ref{eg}, bottom right). However it is still a matter of investigation as to why some merging clusters host radio halos and relics while others do not.

For the reacceleration mechanisms to work, a seed population of relativisitic electrons is needed as the efficiencies of acceleration are low \citep[e. g.][]{mar05,kan11,pin17}. These seeds may be due to the radio galaxies in clusters and the secondary electrons. Low frequency observations are crucial in order to trace the seed population as it is expected to be aged synchrotron plasma with steep spectra.

In this review I will focus on how low frequency observations have helped to get insights into the occurrence statistics of radio halos and relics, led to the discoveries of such diffuse sources in low mass clusters and are helping to trace the seed populations of relativistic electrons.

\section{Radio surveys of galaxy clusters}
\begin{figure}[t]
\begin{center}
 \includegraphics[trim= 2.0cm 2.0cm 1.0cm 3.0cm,clip,width=4.5in]{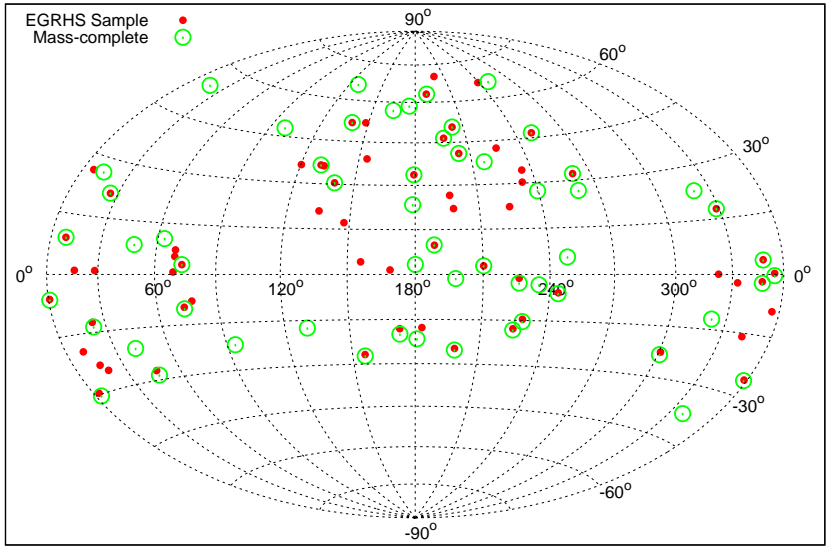} 
 \caption{The largest uniform surveys of galaxy clusters aimed towards searching for diffuse radio emission from the intra-cluster medium are shown on the celestial sphere. The clusters surveyed in the Extended GMRT Radio Halo Sample (\citet{ven07,ven08,kal13,kal15}) are  marked with red filled points. The clusters that form the mass-complete extension \citep{cuc15} of the EGRHS sample are shown in green open points.}
   \label{egrhs}
\end{center}
\end{figure}

Typically the search for diffuse radio emission has been done in all sky surveys such as the NRAO VLA Sky Survey (NVSS), WENSS and more recently in the GLEAM survey. Although this approach can result in discoveries, it is not useful to quantify the non-detections. For that purpose tailored surveys such as the GMRT Radio Halo Survey \citep{ven07,ven08} are crucial. A short summary of the recent and ongoing surveys is provided below.\\
{\underline{\it The Extended GMRT Radio Halo Survey}}\\
The EGRHS is an extension of the GRHS and together have surveyed a total of 64 clusters selected from the REFLEX and eBCS catalogues of clusters \citep{kal13,kal15}. 
These were the clusters in the redshift range 0.2 - 0.4 and brighter than $5\times10^{44}$ erg s$^{-1}$ in X-ray luminosity and were above the declination of $-31^\circ$ (Fig.~\ref{egrhs}). Radio halos were found in $22\%$ and relics in $5\%$ of the sample. Upper limits using the method of injection of models \citep{ven08} were reported on a total of 31 clusters. Using X-ray morphology estimators, strong evidence for the connection between cluster merger and occurrence of radio halos was found. This is the first largest uniformly surveyed sample of galaxy clusters and formed the basis for further exploration of radio power - cluster mass scaling and a study of the Brightest Cluster Galaxies \citep[e. g.][]{cas13,kal15b}.
\\
{\underline{\it Mass complete sample: Ongoing survey}}\\
The detections of a large number of galaxy clusters by the Planck satellite using the Sunyaev-Zel'dovich effect resulted in a sample of clusters with well constrained total masses \citep{2014A&A...571A..29P}. A sample of galaxy clusters with mass-completeness $\sim80\%$ in the redshift range 0.08 - 0.33 with masses, $>6\times10^{14}$ M$_{\odot}$ was constructed and explored occurrence rates of radio halos \citep{cuc15}. The sample is shown in Fig.~\ref{egrhs}. The fraction of radio halos in high and low mass sub-samples divided at $8\times10^{14}$ M$_{\odot}$ was found to be $60-80\%$ and $20-30\%$, respectively. Two underluminous radio halos have been found in this sample \citep{cuc18} and the analysis of the remaining sample is ongoing and will provide the statistics of occurrence in the coming years.

\section{Low mass, merging galaxy clusters}
\begin{figure}[t]
\begin{center}
 \includegraphics[trim= 11.3cm 5.5cm 9.0cm 5.5cm,clip,height=2.2in]{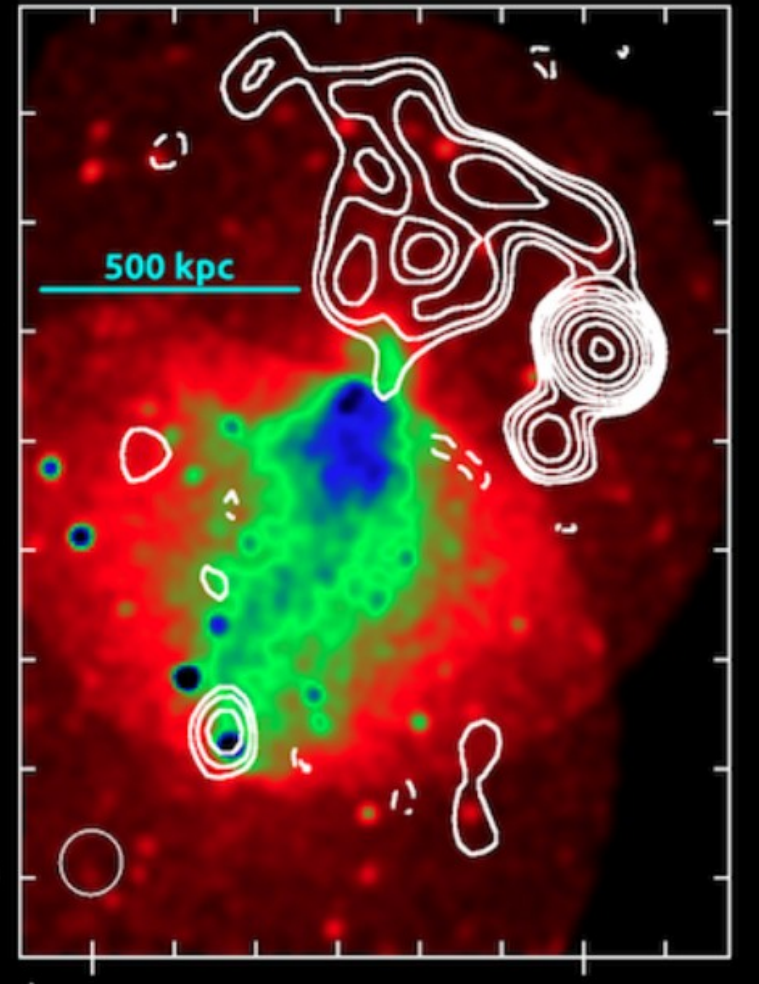} 
  \includegraphics[trim= 2cm 2cm 3.5cm 2cm,clip,height=2.2in]{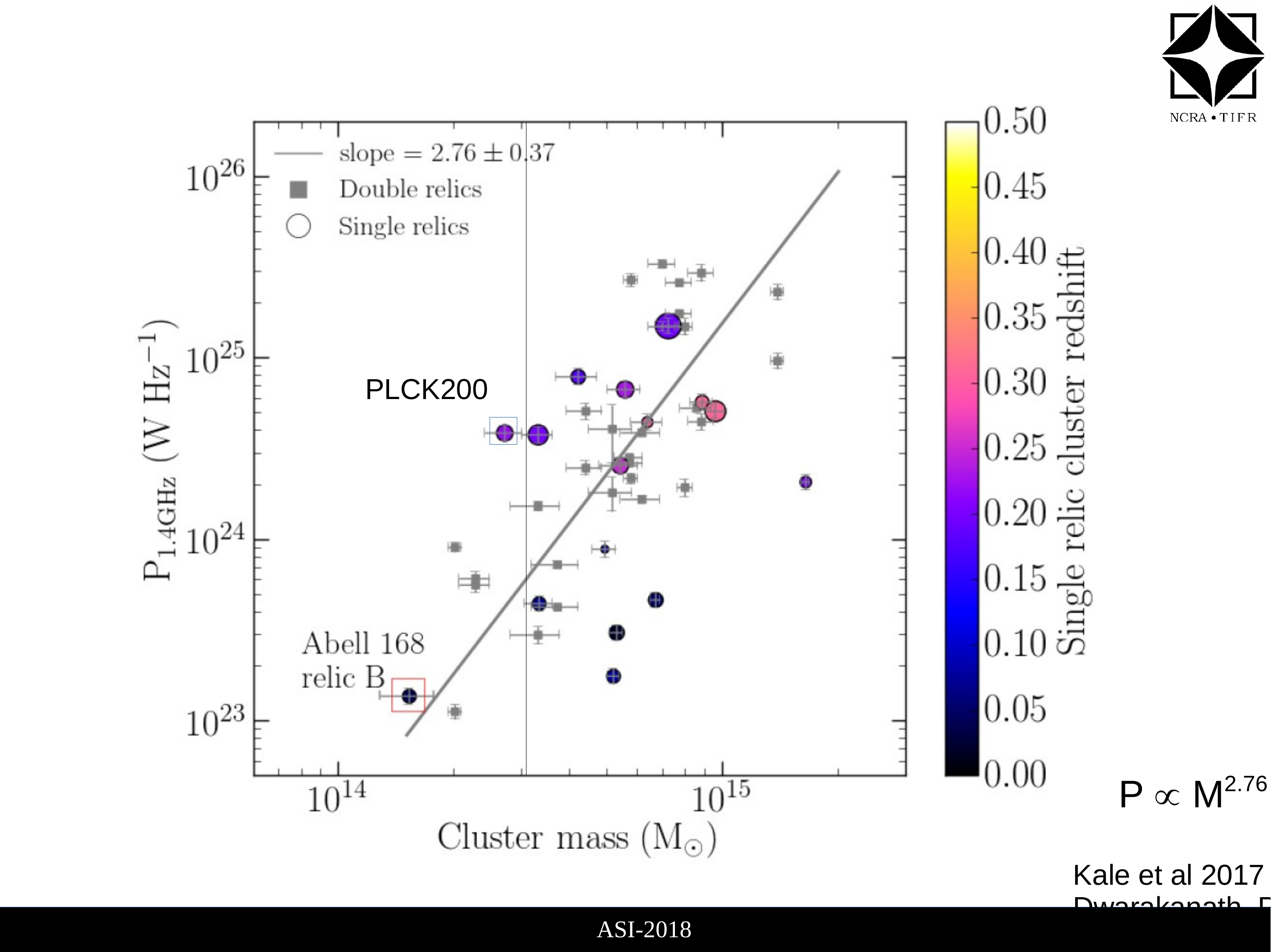} 
 \caption{Left:- The single relic cluster with lowest host cluster mass discovered from the GLEAM survey. Right:- The radio power versus cluster mass plane for radio relics. The double radio relics (grey squares) and the single relics (circles scaled according the relic size and coloured according to the redshift shown in the color bar) are shown. The vertical line is plotted to highlight the two lowest mass clusters with single radio relics labelled as ''PLCK200" and ''Abell168 relic B". The best fit line for the scaling relation plotted as the grey solid line. Images in both the panels are adapted from \citet{dwa18}.}
   \label{lowmass}
\end{center}
\end{figure}
The EGRHS and the mass-complete samples have focused mainly on the most massive clusters. The low mass clusters that show merging signatures are promising sites to constrain the properties of shocks and turbulence. A radio relic was recently used to trace a low mass cluster \citep{gas17}.

We carried out a search for diffuse radio emission of extent $>500$ kpc towards the newly detected clusters in the Early SZ Planck catalogue \citep{2015A&A...581A..14P}. This resulted in the discovery of a radio halo in the cluster PLCKG171.9-40.7 \citep{gia13} and  a single radio relic in the cluster PLCK G200.9-28.2 \citep{kal17}.
The single relic cluster with the mass of $2.7\times10^{14}$ M$_{\odot}$ was the lowest mass host to a single relic known at that time. We found an even lower mass cluster with a single relic using Murchison Widefield Array (MWA,\citet{2013PASA...30....7T}) which has the best existing short baseline coverage at low frequencies ($<200$ MHz) in the southern hemisphere.
The cross matching of the galaxy cluster catalogues with the 200 MHz GaLactic and Extra-galactic All-sky MWA survey (GLEAM, \citet{2017MNRAS.464.1146H}) led to the discovery of a radio relic in the cluster Abell 168 \citep{dwa18}. The confirmation of the relic was done with the GMRT (325 and 610 MHz) and Very Large Array (1 - 2 GHz).
The single arc-like relic at the periphery of this cluster has a smaller ring-like radio source in its wake. Based on the curved spectrum of this source we termed this as a remnant relic that may have been responsible for providing seed relativistic electrons in the outer relic. This is a unique case where the reaccelerated plasma at the shock and the possible seed plasma are detected.

\section{Broadband studies with the Upgraded GMRT: characterising the seeds}
The remnants of radio galaxies in galaxy clusters can be important sources of seed relativistic electrons. Galaxy clusters have been found to be host to diffuse sources that are filamentary or ring-like with sizes of 100 - 200 kpc and with steep spectra \citep{sle01}. A number of such sources have been recently discovered using deep radio observations \citep[e. g.][]{shim15,shim16}. Such remnants can be revived by adiabatic compression due to shocks \citep{ens01} or by gentle reacceleration \citep{wee17,gasp17}. In order to characterise the seed population, their spectra need to be studied. The spectra can often be curved 
and can vary across the extent of the source \citep[e. g. 1 - 2 GHz study of A2256 relics][]{2014ApJ...794...24O}. 

\begin{figure}[t]
\begin{center}
 \includegraphics[trim= 2.0cm 4.0cm 1.0cm 3.0cm,clip,width=5in]{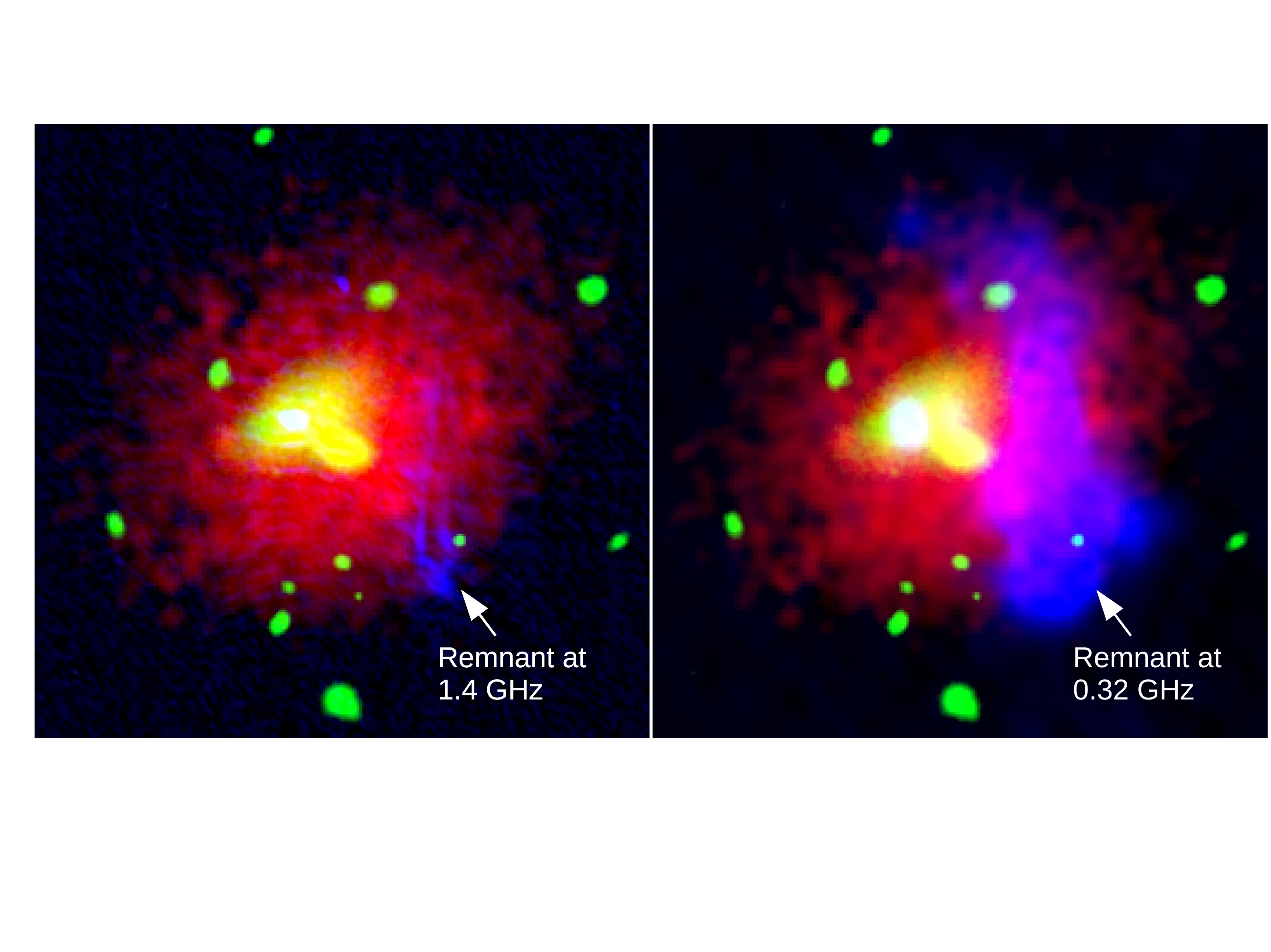}
 \caption{The cluster Abell 4038 in optical (green), X-ray (red) and radio bands (blue) is shown in both the panels. The 1.4 GHz image is shown in the left and the 0.32 GHz image is shown in the right panel. The low frequencies trace the steep spectrum emission that is much more extended compared to the higher frequencies.}
   \label{remnant}
\end{center}
\end{figure}

The Upgraded GMRT (uGMRT) is now operational and provides observations with near seamless coverage between 0.12 - 1.4 GHz with instantaneous bandwidths of up to 400 MHz \citep{Gupta2017}. The observing bands have been released on shared risk basis starting mid-2016. The uv-coverage has improved due to the enhanced bandwidth and  extended sources can be imaged up to sizes twice of what were possible to be imaged with the legacy GMRT \citep{2017ExA....44..165D}. The radio relic in Abell 4038 is a prototype of steep spectrum relics which require low frequency observations to reveal their extent (Fig.~\ref{remnant}). Our GMRT study led us to the conclusion at it may be an adiabatically compressed remnant of the radio galaxy hosted by the Brightest Cluster Galaxy \citep{kaldwa12}. We have mapped the spectrum across the extent of the radio relic in the cluster Abell 4038 \citep{kaldwa12} with the uGMRT in the band 300 - 500 MHz and 1050 - 1450 MHz and found variations of the spectral curvature across the extent of the relic \citep[][]{2018MNRAS.tmp.2116K}. Images with rms $70\mu$Jy beam$^{-1}$ at 0.4 GHz and $30 \mu$Jy beam$^{-1}$ 1.26 GHz were made and used to study the variation of the spectral index and its curvature across the relic. Between 0.3 - 1.4 GHz, the spectral curvatures range between 0.5 - 1.6, with the highest curvatures in a curved region of the relic skirting an arc-like feature seen in the X-ray image. The variations thus are connected to the local physical conditions and possibly indicate a change in magnetic field across the source as well.


\section{Conclusions}
Radio halos and relics are direct probes of the non-thermal components in the ICM. Sensitive low frequency observations possible with the current and upcoming interferometers are well suited to reveal the spectra and morphologies of these sources to constrain the theoretical models. In recent years the role of reacceleration by turbulence and shocks in the formation of radio halos and relics has gained considerable support. The availability of seed relativistic electrons so that the reacceleration mechanisms are efficient may be one of the crucial factor dividing merging clusters with and without radio halos and relics. The uniform surveys of galaxy clusters, namely the EGRHS and the ongoing survey mass-complete sample are providing the occurrence statistics of radio halos and relics. The unexplored regime of low mass merging clusters is becoming accessible due to the sensitive low frequency telescopes such as the MWA (see M. Johnston-Hollitt, this volume). The seed relativistic electron population is likely going to be steep spectrum and thus their characterisation relies on sensitive measurements in sub-GHz frequency ranges. The uGMRT is operational and we have carried out the first study of the spectrum of a remnant radio galaxy in the cluster A4038.
The prospects to study radio halos and relics with the uGMRT are promising.

\acknowledgements
RK would like to thank the organisers for the invitation to give a talk at this excellent conference. RK acknowledges the support from the DST INSPIRE Faculty Award by the Government of India. We thank the staff of the GMRT that made these observations possible. GMRT is run by the National Centre for Radio Astrophysics of the Tata Institute of Fundamental Research. This research has made use of 
the NASA/IPAC Extragalactic Database (NED) which is operated by the Jet 
Propulsion Laboratory, California Institute of Technology, under contract with 
the National Aeronautics and Space Administration. This research made use of 
Astropy, a community-developed core Python package for Astronomy (Astropy 
Collaboration, 2018). The scientific results reported in this article are based 
in part on data obtained from the Chandra Data Archive. The National Radio Astronomy Observatory is a facility of the National Science Foundation operated under cooperative agreement by Associated Universities, Inc. Based on observations obtained with XMM-Newton, an ESA science mission with instruments and contributions directly funded by ESA Member States and NASA

\bibliographystyle{apj}
\bibliography{mybib_1,mn-jour}

\end{document}